\documentclass[]{jkas} 
\usepackage{ifsym}
\usepackage{amssymb, amsmath}

\def\year{2025} 
\def\volume{56} 
\def\issue{1} 
\def\beginpage{1} 
\def\received{October 12, 2022} 
\def\accepted{December 20, 2022} 
\def\published{January ??, 2023} 

\date{Received \received; Accepted \accepted; Published \published}

\def\sun{\odot}
\newcommand{\qdist}[1]{\ifmmode\langle#1\rangle\else\textlangle#1\textrangle\fi}
\def\arcsec{\hbox{$^{\prime\prime}$}} 

\newcommand{\tred}[1]{{\textbf{#1}}}

\title{%
SDSS J134313.15$+$364457.5: Forming Compact Elliptical through the Merger
}

\author[1]{Daya Nidhi Chhatkuli\thanks{The original author of this document.}}{0000-0002-8040-6902}
\author[2,3$\star$]{Sanjaya Paudel}{0000-0003-2922-6866}
\author[4]{Binil Aryal}{0000-0002-1253-0741}
\author[5]{Binod Adhikari}{0000-0002-2799-9112}
\author[6]{Nau Raj Pokhrel}{0000-0003-2569-8129}
\author[7]{Rajendra Adhikari}{}

\affil[1]{Department of Physics, Tri-Chandra Multiple Campus, Tribhuvan University, Nepal}
\affil[2]{Department of Astronomy, Yonsei University, Seoul, 03722, Republic of Korea}
\affil[3]{Nepal Astronomical Society, Kathmandu, Nepal}
\affil[4]{Central Department of Physics, Tribhuvan University, Kirtipur, Nepal}
\affil[5]{Patan Multiple Campus, Tribhuvan University, Lalitpur, Nepal}
\affil[6]{Department of Physics and Astronomy, The University of Tennessee, 1408 Circle Drive, Knoxville, TN 37996, USA}
\affil[7]{Amrit Science Campus, Tribhuvan University, Kathmandu, Nepal}

\def\corrauthor{%
S. Paudel, \email{sanjpaudel@gmail.com}
}

\def\runningauthor{%
Chhatkuli et al.
}

\def\runningtitle{%
Forming Compact Elliptical through the Merger
}

\def\keywords{%
Galaxy: individual, Galaxy: compact, Galaxy: dwarf galaxy, Galaxy: interaction
}

\def\abstracttext{%
Scaling relations are fundamental tools for exploring the morphological properties of galaxies and understanding their formation and evolution. Typically, galaxies follow a scaling relation between mass and size, measured by effective radius. However, a compact class of galaxies exists as outliers from this relation, and the origin of these compact galaxies in the local universe remains unclear. In this study, we investigate the compact dwarf galaxy SDSS J134313.15+364457.5 (J1343+3644), which is the result of a merger.  Our analysis reveals that J1343+3644 has a half-light radius of 482~pc, significantly smaller than typical galaxies with the same brightness ($M_\text{r} = -19.17$ mag). With a high star-formation rate (SFR) of 0.87~M$_{\sun}$ year$^{-1}$, J1343+3644 is expected to evolve into a compact elliptical galaxy in a few million years. J1343+3644 could, therefore, be a progenitor of a compact elliptical galaxy. The phenomenon happened in early universe, where compact galaxies were common.
}

\begin{document}
\jkashead 
\section{Introduction}
 Structural parameters—such as size, concentration, and ellipticity—are essential tools for understanding galaxy formation and evolution. Galaxies follow a continuous scaling relationship between their size and luminosity across a broad range of magnitudes \citep{Graham03, Ferrarese06, Janz08} and such a scaling relation is commonly used as a test-bed for theoretical galaxy formation models. Nonetheless, the extensive applicability of the relationship has ignited discussions, owing to outliers and their non-linear characteristics.\citep{Kormendy85, Kormendy09, Chen10}. Early-formed galaxies are notably more compact than those formed later, evident from the smaller sizes of high-redshift galaxies compared to their local counterparts \citep{Daddi07}. This raises intriguing questions about the formation and evolution of compact galaxies in the early universe, a significant mystery in modern cosmology.

Compact dwarf galaxies deviate significantly from the standard size-magnitude relation \citep{Chilingarian09, Zhang17}, with their formation and evolution poorly constrained. They are classified into star-forming and quiescent types. Star-forming Blue Compact Dwarfs (BCDs) are small, dense systems with intense blue emission from young, massive stellar clusters \citep{Papaderos96,Sung98}. Quiescent compact early-type galaxies (cEs), such as M32, lack active star formation. Ultra-compact dwarfs (UCDs), with sizes comparable to a few times that of globular clusters, represent an extreme subset \citep{Mieske02}.

cEs are frequently found near larger galaxies like M31 (the Andromeda Galaxy). This proximity has led to the hypothesis that tidal stripping by nearby giants may explain their formation \citep{Faber73, Huxor11, Chilingarian15}. However, not all compact dwarfs reside near large galaxies. \citet{Huxor13} discovered one in a field environment, suggesting an alternative origin via mergers of smaller dwarfs. Supporting this, \citet{Paudel14} identified an isolated compact dwarf with clear merging features. In addition, a study on a large sample of cEs of redshift range z $<$ 0.05, \cite{Kim20} suggests that cEs comprise a mixture of galaxies
with two types of origins depending on their local environment. This view has also been supported by the fact that cE also follows the scaling relation between Black Hole mass and and mass-bulge luminosity \citep{Rey21,Paudel16}.

Mergers between massive galaxies play a crucial role in galaxy evolution, driving morphological transformations from star-forming disk galaxies to non-star-forming ellipticals \citep{Toomre72, Naab09}. However, the significance of mergers at low masses has been debated, as low-mass galaxies are more influenced by their local environment. Recent observations show that dwarf-dwarf mergers are not uncommon \citep{Paudel18,Paudel14,Stierwalt15}. Features such as stellar shells, kinematically decoupled cores, and tidal streams in dwarf galaxies are frequently observed, often linked to these mergers \citep{Geha05, Rich12, Penny12, Toloba14}.

Observational and theoretical evidence suggests that compact galaxies follow diverse evolutionary paths, often involving central starbursts or tidal stripping. Interactions like dwarf-dwarf mergers or gas inflows can trigger intense gas collapse and dissipation, leading to bursts of star formation in these low-mass systems \citep{Papaderos96, Nusser05, Bournaud09, Elmegreen12, Zhang20a, Nusser22}.

In this study, we provide an analysis of the structural parameters for the compact merging dwarf galaxy SDSS J134313.15$+$364457.5 (hereafter J1343$+$3644). We derive its size and place it on size-magnitude relation, and explore its possible evolutionary scenario. We conduct a comparative study between morphological and star-formation properties.

\section{Data Analysis}
\subsection{Sample Selection}
J1343 $+$ 3644 is taken from a catalog of merging dwarf galaxies prepared by Paudel et al 2018. It has a sky position RA (J2000) $=$ 13$^{\text{h}}$ 43$^{\text{m}}$ 13.13$^{\text{s}}$ and Dec. (J2000) $= +\,36^\circ\,44'\,57.48''$, and a redshift of $z = 0.0197$. In the catalog, this galaxy is featured as the galaxy having a tidal stream. It is found in a field environment where no nearby companion is identified within the 500~kpc sky projected radius with a line of sight relative radial velocity range of $\pm500$~km~s$^{-1}$.

\begin{table}[h]
\caption{Basic physical parameter of J1343$+$3644}
\centering
\begin{tabular}{l|l}
\hline
Parameter & Value \\
\hline
RA  & 205.80470$^{\circ}$ \\
Dec & 36.74930$^{\circ}$  \\
$z$ & 0.0197 \\
$M_{\text{r}}$  & $-$19.17 mag \\
$R_{h}$  & 482 pc \\
$g-r$ & 0.38 mag \\
SFR & 0.87 M$_{\sun}$ year$^{-1}$ \\
$12+\log$[O/H]   & 8.44 dex \\
HI mass & 7.9 $\times$ 10$^{9}$ M$_{\sun}$\\
\hline
\end{tabular}
\label{par1}
\\
The SFR is calculated from H$_{\alpha}$ line flux using the relation from \citet{Kennicutt98}. Gas-phase metallicity (12 + log[O/H]) is determined from the H$_{\alpha}$
/[NII] flux ratio, calibrated by \citet{Marino13}. Neutral hydrogen (HI) mass is sourced from the FASHI catalog \citep{Zhang24}.
\end{table}

\begin{figure}[h]
\centering
\includegraphics[width = 8.5cm]{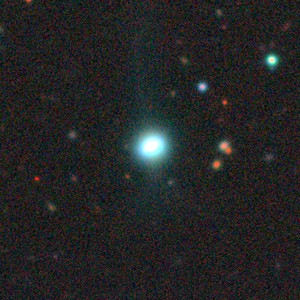}
\caption{Optical images of J1343$+$3644. The field of view of the image is $1\times1$~arcmin, where north is the top and east is left. The image is obtained from the Legacy survey sky-server.}
\label{main}
\end{figure}

Figure~\ref{main} shows the optical image of J1343$+$3644 in which an elongated low surface brightness tidal tail along the north direction can be seen, while the overall shape of the galaxy is fairly round. The color image is a cropped section from the Legacy sky-server images \citep{Dey19}, prepared by merging images from the  $g-$, $r-$, and $z-$ bands, which was observed by the Bok 2.3-m telescope (BASS). The field-of-view of the image is $1'\times1'$. The color reveals that the tidal tail component has an older, non-star-forming stellar population compared to the galaxy's main body.

The basic photometric properties of J1343$+$3644 are listed in Table 1. It has a $r-$band absolute magnitude $M_{\text{r}} = -19.17$~mag and an overall $g-r$ color index of 0.38~mag, with a star-formation rate of 0.87~M$_{\sun}$ year$^{-1}$.  The measured half-light radius is 482~pc.

\subsection{Imaging analysis}
We used the SDSS imaging data to perform image analysis \citep{Ahn12}. The $r-$band image is usually taken to do surface photometry because it has a higher signal-to-noise ratio than the other bands. We fetched the SDSS DR12 imaging data, which has relatively better sky-background subtraction compared to the previous data release. We, however, subtracted the sky background, calculating a median value from 10 randomly selected boxes of $10\times10$ around the galaxy. 

\begin{figure}[h]
\includegraphics[width=8cm]{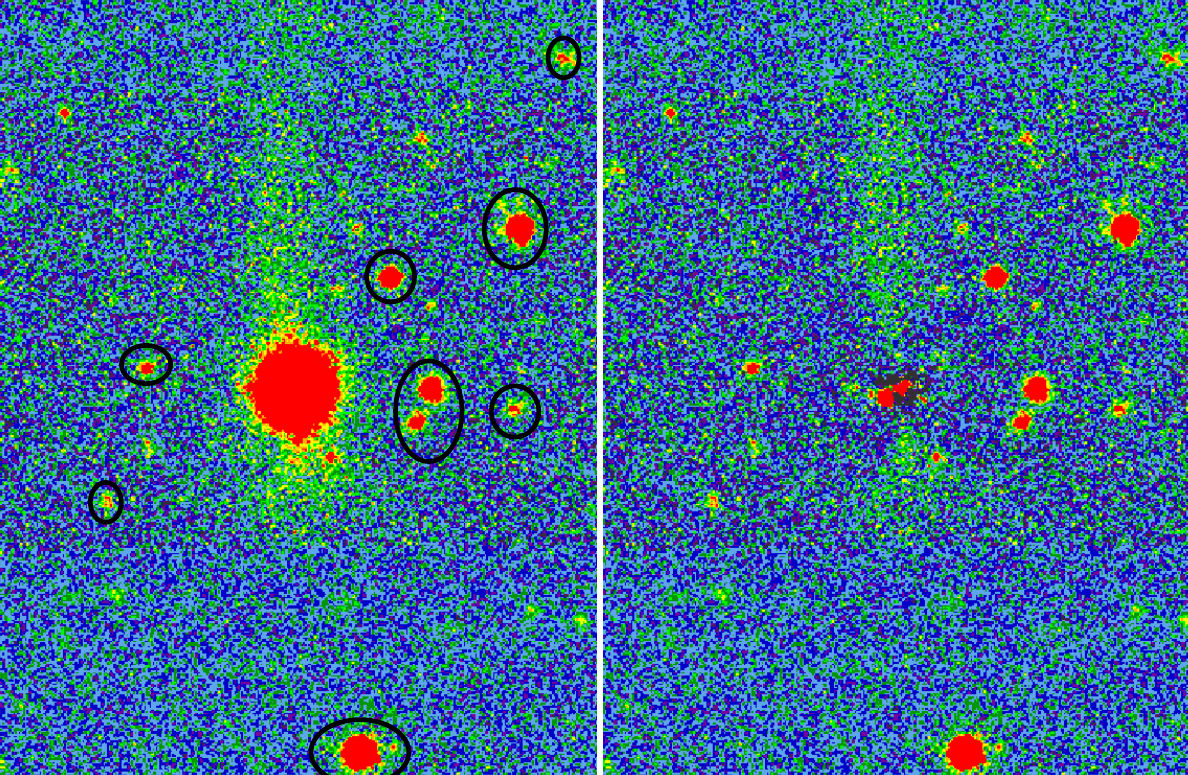}
\includegraphics[width=8.5cm]{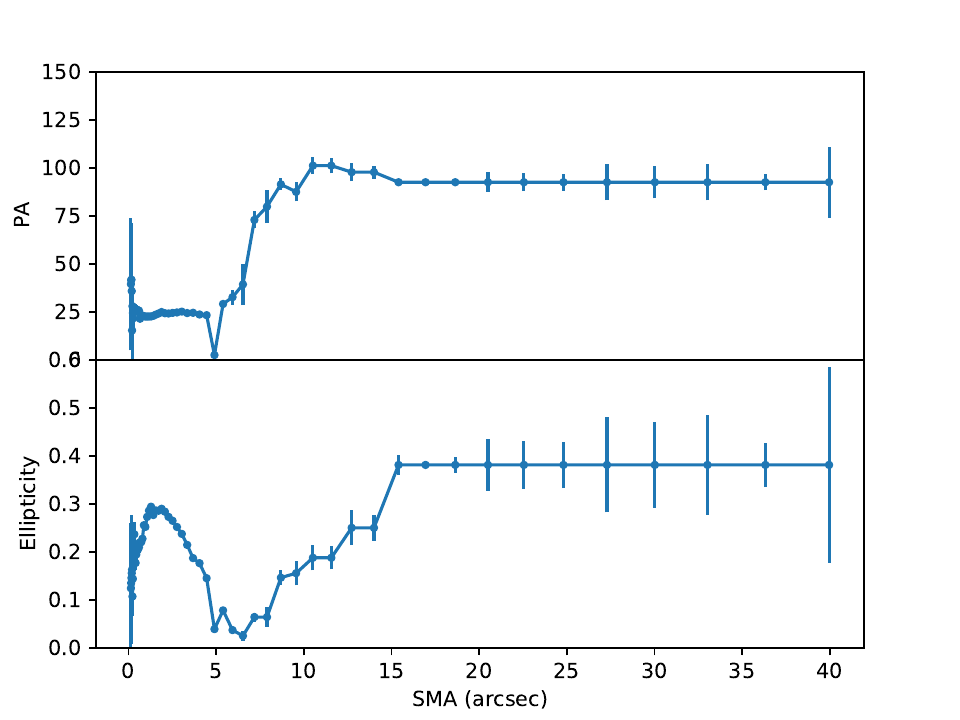}
\caption{Upper: We show the SDSS $r-$band image and its residual image after subtracting the best-fit ellipse model. \tred{The black ellipse area represents the masked region during the ellipse fit. }
Lower: we show variation of ellipse parameter (Position angle on top and Ellipticity on bottom) along the major axis.} 
\label{isopro}
\end{figure}

The Image Reduction and Analysis Facility (IRAF) task ellipse was used to extract the galaxy’s major-axis light profile. IRAF is a widely used suite of software tools for astronomical data analysis. The $ellipse$ task fits elliptical isophotes to galaxy images and computes parameters such as eccentricity, intensity, radial distance, position angle, and ellipticity. Based on the methodology outlined in \cite{Jedrzejewski87}, it performs azimuthal averaging of intensity $I(\phi)$ along the elliptical contour, using initial estimates for the isophote center, ellipticity, and semi-major axis position angle.

We show the optical $r-$band image of J1343$+$3644 in the upper left panel of Fig. \ref{isopro}.  To perform surface photometry, we first masked the unrelated foreground and background objects through manual identification, see black ellipse. The galaxy center was determined using the $imcentre$ task in IRAF, which computes the centroid of the specified image section. During ellipse fitting, the position angle and ellipticity were allowed to vary freely, while the semi-major axis was incremented logarithmically. The ellipse center was constrained to deviate no more than 3 pixels between consecutive isophotes. We present the residual image obtained by subtracting the best-fit ellipse model from the original image in the upper-right panel, where we can see a prominent tidal tail of the galaxy. 

The lower panel of Figure~\ref{isopro} illustrates the changes in position angle (PA) and ellipticity ($\epsilon$) along the major axis. The position angle exhibits rapid variations in the inner region and stabilizes beyond a radius of ten arc seconds. Meanwhile, the ellipticity displays similar trends, remaining relatively constant in the central region and steadily increasing towards the outer radius.

\begin{figure}
\centering
\includegraphics[width=8cm]{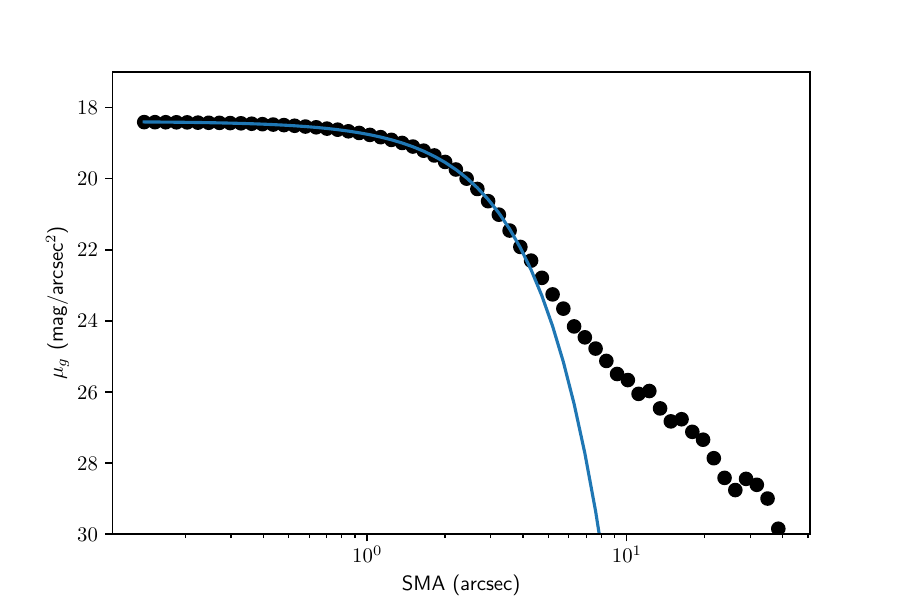}
\caption{We show the one-dimensional surface brightness profile of J1343$+$3644. The black dot represents the observed point and the blue line represents best fitted S\'ersic model.} 
\label{serfit}
\end{figure}

\subsection{Size of the Galaxy}
Galaxies lack a distinct endpoint, making the measurement of their size a non-trivial task. Typically, two approaches are employed to determine galaxy size: parametric and non-parametric. In the parametric approach, the galaxy's light profile is initially approximated using a parametric function, such as exponential or de Vaucouleurs's. It's worth noting that both exponential and de Vaucouleurs's functions are specific instances of the S'ersic function \citep{Sersic68}, which is generally defined as:

\begin{equation}
I(R) = I_{\rm e}\exp\left\{-b_n\left[\left( \frac{R}{R_{\rm e}}\right) ^{1/n} -1\right]\right\}, \label{EqSer}
\end{equation}
Here, $I_{\rm e}$ represents the intensity of the light profile at the effective radius $R_{\rm e}$, and $n$ characterizes the 'shape' of the profile. The term $b_n$ is a function of $n$ chosen to guarantee that the radius $R_{\rm e}$ encompasses half of the profile's total luminosity. Typically, for early-type galaxies known as de Vaucouleurs's, we use $n = 4$, while for late-type galaxies, we opt for $n = 1$, corresponding to an exponential profile.

In Figure~\ref{serfit}, we have modeled the galaxy light profile with a S\'ersic function. It is clear that we are able to fit only the inner part of the galaxy, and the outer tidal component is left. The derived Sersic index and effective radius are 0.92 and 470~pc, respectively. Since the outer part of J1343$+$3644 possesses a tidal feature and it is well expected that no S\'ersic model provides the best fit to the overall light profile of such a galaxy. We, therefore, use a non-parametric approach to derive an overall galaxy structure parameter.

Nonparametric techniques determine a galaxy's size without relying on any analytical function. A straightforward approach is the Petrosian method. Initially, we establish an empirical radius from the observed intensity profile, referred to as the Petrosian radius $a_{\text{p}}$ (\citealt{Blanton01,Petrosian76}). The calculation involves determining the Petrosian index, defined as the ratio of the surface brightness at a specific distance $R$ to the mean surface brightness within $R$, i.e.,

\begin{equation}
n(R) = \frac{\mu(R)}{\qdist{\mu(R)}}
\end{equation}
Here, $\mu(R)$ represents the surface brightness at radius $R$, and $\qdist{\mu(R)}$ is the average surface brightness within that radius. The Petrosian radius is defined at $n = 0.2$. It is then assumed that the galaxy extends up to a maximum of $2a_{\text{p}}$, and the total flux of the galaxy is determined by the flux measured within a $2a_{\text{p}}$ aperture. Once we know the total flux, we can derive a half of the total flux at a galactocentric distance, commonly known as a half-light radius. We derive a half-light radius for the galaxy J1343$+$3644, which is 1.12~arcsec (482 pc).
 
\begin{figure}[h]
\centering
\includegraphics[width=8.5cm]{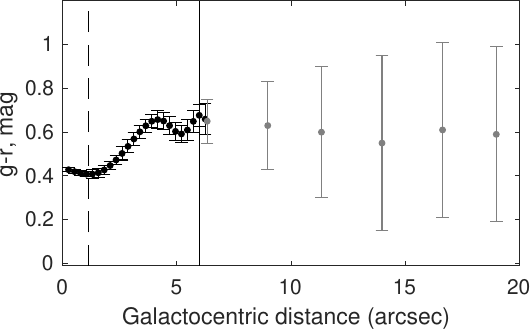}
\caption{$g-r$ colour gradient along the major axis. The position of the dashed line represents the half-light radius and the solid line represents the position where the tidal tail becomes prominent. The color profile beyond 6~arcsec (depicted in grey) is obtained by binning pixels placed outward along the major axis.}
\label{crgrad}
\end{figure}

In Figure~\ref{crgrad}, we show $g-r$ color gradient, which is derived from the azimuthally averaged light profiles of the galaxy's main body. For the tidal tail, we use a manual selection of region box of $10\times10$ pixel; see the gray symbol. The color profile reveals that the inner part of the galaxy is significantly bluer than the outer tidal tail. This difference in color hints that J1343$+$3644 may have accreted a non-star-forming dwarf galaxy. 

\subsection{Spectroscopic Measurement}
\begin{figure}[h]
\centering
\includegraphics[width=8.5cm]{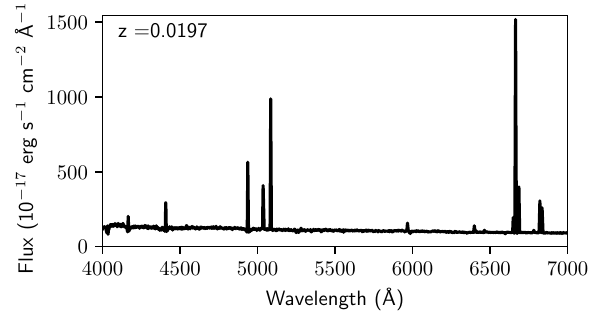}
\caption{Optical spectrum of J1343$+$3644. We can see prominent emissions in the Balmer lines.}
\label{spec}
\end{figure}

We retrieved the optical spectrum of J1343$+$3644 from the SDSS data archive. A signal-to-noise ratio of the spectrum is good ($>$100).  The spectrum is observed with a fiber spectrograph of  3$\arcsec$ diameter. For the distance of J1343$+$3644, i.e  83~Mpc, the central 3$\arcsec$ represents a 1.22~kpc core region which is significantly larger than galaxy size. Therefore we assume that the SDSS fiber spectrum well samples the galaxy core region. 

The SDSS spectra presented in Figure~\ref{spec} reveal prominent emission lines, particularly the Balmer lines, resembling a typical spectrum of a star-forming HII region. To measure emission line flux, we subtract the stellar absorptions  \citep{Tremonti04,Sarzi06}. We have used the IRAF {\it splot} task to measure the emission line flux by fitting a Gaussian profile. The task also measures the FWHM and equivalent width simultaneously. 

The Oxygen abundances ($12+\log$[O/H]) are derived from the NII method, which uses the ratio of H$_\alpha$ and [NII] emission line fluxes. For this, we used a calibration provided by \citet{Marino13} and obtained $12+\log$[O/H] = 8.44~dex. 

Using the H$_\alpha$/H$_\beta$ emission-line ratios, we derived the internal extinction coefficient, $E(B-V)$ \citep{Cardelli89}. The calculated extinction coefficient is 0.7~mag. Subsequently, we determined the star formation rate (SFR) by correcting the H$_\alpha$ emission line flux for extinction, employing the calibration provided by \citet{Kennicutt98}. The SFR of the galaxy is obtained to be 0.87~M$_{\sun}$ year$^{-1}$. 

The measured equivalent width of H$_\alpha$ emission line is 89~\AA. Starburst99 model \citep{Leitherer99} for an instantaneous burst of star formation gives star formation age order of 100~Myr. 

\section{Conclusion and Discussion}
\begin{figure}[h]
\centering
\includegraphics[width=8.5cm]{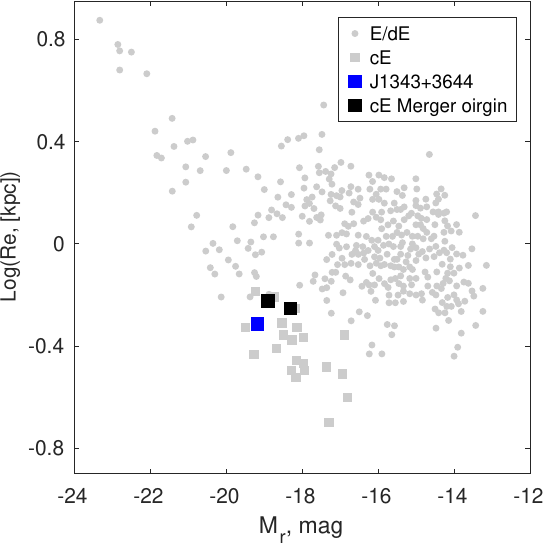}
\caption{The size-magnitude relation of early-type galaxies, spanning from dwarf ellipticals to giant ellipticals, including the outlier cE. The plot is reproduced from \citep{Paudel14}, where Es/dEs are shown in dots, and the square represents cEs. The  J1343+3644 is marked with blue, and for comparison, the positions of previously identified merger-origin cEs are indicated with black. }
\label{smag}
\end{figure}

This study focuses on the compact merging dwarf galaxy J1343$+$3644 and provides a comprehensive analysis of its morphological parameters. The derived size of J1343$+$3644 is 1.12 arcseconds, which corresponds to a physical radius of 482 pc. Its position on the size-magnitude diagram, shown in Figure \ref{smag}, places it significantly below the average size-magnitude relation defined by normal galaxies, highlighting its compact nature. 

We compare J1343$+$3644 with previously identified cEs believed to have formed through mergers \citep{Huxor13, Paudel14}, including a reference sample from \citep{Chilingarian15}. Notably, J1343$+$3644 appears to be the most compact cE among those with a merger origin.

\subsection{Future evolution of J1343+3644}

\begin{figure}[h]
\centering
\includegraphics[width=8.5cm]{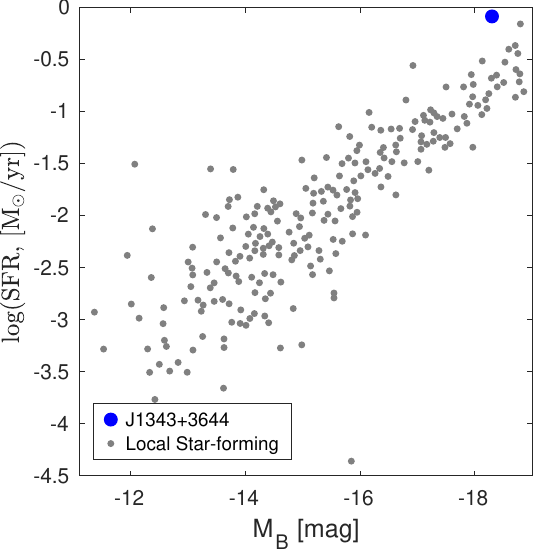}
\caption{SFR-B-band magnitude relation of star-forming galaxies. J1343$+$3644 is denoted by blue symbols. The gray symbols represent a comparison sample obtained from \cite{Lee09}.}
\label{gfig}
\end{figure}

Figure~\ref{gfig} displays the correlation between the B-band absolute magnitude and the logarithm of star formation rate ($\log$(SFR)) for star-forming galaxies. The star-formation rate of J1343+3644 is derived from the H$_{\alpha}$ flux of the SDSS spectrum. The B-band magnitude of our sample galaxies is calculated from the SDSS $g-$band magnitude using the color transformation equation provided by the SDSS webpage\footnote{http://www.sdss3.org/dr8/algorithms/sdssUBVRITransform.php} [$B = g + 0.39\times(g-r) + 0.21$]. The comparison sample in gray dots is taken from \citet{Lee09}. This study offers a statistical analysis of star formation activity in star-forming galaxies within the local volume, utilizing a volume-limited sample. Interestingly, J1343 $+$ 3644 is a significant outlier from the main-sequence relation defined by normal star-forming galaxies, which means J1343+3644 is forming stars significantly at a higher rate than normal star-forming galaxies. Indeed active star-formation is a sign of interaction which was previously identified by \citep{Stierwalt15}

The FAST all-sky HI survey (FASHI) survey has estimated the total gas mass to be 7.9 $\times$ 10$^{9}$ M$_{\odot}$ \citep{Zhang24}. Assuming a conversion efficiency of 20 percent from HI mass to stellar mass \citep{Kravtsov18}, it is anticipated that star formation can only be sustained for a few hundred mega-years.
What changes can be expected in the evolution of J1343$+$3644 following the cessation of star-formation activity? It is probable that the sizes of star-forming galaxies will increase as dynamical and stellar population evolution takes place, resulting in a decline in surface brightness \citep{Wellons15, Wellons16}. J1343$+$3644 is significantly compact compared to a typical galaxy. Even if it will expand in the future by doubling its size, it will still be a compact galaxy. The SDSS spectrum also shows the presence of a higher-order Balmer absorption feature, suggesting that the star-formation activity is recent and has a notable portion of the old stellar population. Since we see that the tidal part is relatively redder than the galaxy's main body, the satellite accretion could have contributed to the old stellar population. The event eventually perturbed the primary gaseous disk, leading to a burst of star formation at the center. 

\citet{Zhang20a,Zhang20b} studied dwarf-dwarf merger and the formation of a compact star-forming dwarf galaxy (VCC\,848) in the Virgo cluster. A JVLA HI emission line map and optical imaging combined reveal an extended gaseous distribution and central star-burst, though VCC\,848 has less than 30\% of gas concentration in the center.  A suite of $N$-body hydrodynamical simulations results suggest the timing of the merger about a Giga year ago. Contrary to VCC\,848 environments, J1343$+$3644 is located in a low, dense environment. 

Early simulations suggest that galaxy mergers induce substantial gas inflows toward the center, sparking intense starbursts that lead to core-like structures \citep{Mihos94}. This effect is pronounced in bulge-less systems, such as star-forming dwarf galaxies, where the absence of a bulge facilitates the development of non-axisymmetric structures (e.g., bars) that channel gas inward. In contrast, a bulge can stabilize a galaxy, suppressing these inflows. The observed central starburst in our merging compact galaxy aligns well with these simulation predictions, highlighting the role of gas dynamics in shaping core formation during mergers. Understanding these processes offers clues about galaxy assembly in the early universe and the role of mergers in the transition from star-forming to non-star-forming, more stable elliptical galaxies. 

The cessation of star formation in J1343$+$3644 is typically observed over a timescale of several hundred million years. Following the cessation of star formation, the system may transit into quiescent states, ultimately evolving into compact elliptical galaxies.


\acknowledgments

DNC expresses sincere gratitude to the University Grants Commission of Nepal for its financial support. SP acknowledges support from the Mid-career Researcher Program (RS-2023-00208957 and RS-2024-00344283, respectively) through Korea's National Research Foundation (NRF).

This study is based on the archival images and spectra from the Sloan Digital Sky Survey (the full acknowledgment can be found at https://www.sdss.org/collaboration/acknowledgements).




\begin{thebibliography}{}
\expandafter\ifx\csname natexlab\endcsname\relax\def\natexlab#1{#1}\fi
\providecommand{\url}[1]{\href{#1}{#1}}
\providecommand{\dodoi}[1]{doi:~\href{http://doi.org/#1}{\nolinkurl{#1}}}
\providecommand{\doeprint}[1]{\href{http://ascl.net/#1}{\nolinkurl{http://ascl.net/#1}}}
\providecommand{\doarXiv}[1]{\href{https://arxiv.org/abs/#1}{\nolinkurl{https://arxiv.org/abs/#1}}}
\providecommand{\dodoilink}[2]{\href{http://doi.org/#1}{#2}}
\providecommand{\doadslink}[2]{\href{#1}{#2}}

\bibitem[{{Ahn} {et~al.}(2012){Ahn}, {Alexandroff}, {Allende Prieto},
  {Anderson}, {Anderton}, {Andrews}, {Aubourg}, {Bailey}, {Balbinot}, {Barnes},
  \& et~al.}]{Ahn12}
{Ahn}, C.~P., {Alexandroff}, R., {Allende Prieto}, C., {et~al.} 2012, \apjs,
  203, 21

\bibitem[{{Blanton} {et~al.}(2001){Blanton}, {Dalcanton}, {Eisenstein},
  {Loveday}, {Strauss}, {SubbaRao}, {Weinberg}, {Anderson}, {Annis}, {Bahcall},
  {Bernardi}, {Brinkmann}, {Brunner}, {Burles}, {Carey}, {Castander},
  {Connolly}, {Csabai}, {Doi}, {Finkbeiner}, {Friedman}, {Frieman}, {Fukugita},
  {Gunn}, {Hennessy}, {Hindsley}, {Hogg}, {Ichikawa}, {Ivezi{\'c}}, {Kent},
  {Knapp}, {Lamb}, {Leger}, {Long}, {Lupton}, {McKay}, {Meiksin}, {Merelli},
  {Munn}, {Narayanan}, {Newcomb}, {Nichol}, {Okamura}, {Owen}, {Pier}, {Pope},
  {Postman}, {Quinn}, {Rockosi}, {Schlegel}, {Schneider}, {Shimasaku},
  {Siegmund}, {Smee}, {Snir}, {Stoughton}, {Stubbs}, {Szalay}, {Szokoly},
  {Thakar}, {Tremonti}, {Tucker}, {Uomoto}, {Vanden Berk}, {Vogeley},
  {Waddell}, {Yanny}, {Yasuda}, \& {York}}]{Blanton01}
{Blanton}, M.~R., {Dalcanton}, J., {Eisenstein}, D., {et~al.} 2001, \aj, 121,
  2358

\bibitem[{{Bournaud} \& {Elmegreen}(2009)}]{Bournaud09}
{Bournaud}, F., \& {Elmegreen}, B.~G. 2009, \apjl, 694, L158

\bibitem[{{Cardelli} {et~al.}(1989){Cardelli}, {Clayton}, \&
  {Mathis}}]{Cardelli89}
{Cardelli}, J.~A., {Clayton}, G.~C., \& {Mathis}, J.~S. 1989, \apj, 345, 245

\bibitem[{{Chen} {et~al.}(2010){Chen}, {C{\^o}t{\'e}}, {West}, {Peng}, \&
  {Ferrarese}}]{Chen10}
{Chen}, C.-W., {C{\^o}t{\'e}}, P., {West}, A.~A., {Peng}, E.~W., \&
  {Ferrarese}, L. 2010, \apjs, 191, 1

\bibitem[{{Chilingarian} \& {Zolotukhin}(2015)}]{Chilingarian15}
{Chilingarian}, I., \& {Zolotukhin}, I. 2015, Science, 348, 418

\bibitem[{{Chilingarian}(2009)}]{Chilingarian09}
{Chilingarian}, I.~V. 2009, \mnras, 394, 1229

\bibitem[{{Daddi} {et~al.}(2007){Daddi}, {Dickinson}, {Morrison}, {Chary},
  {Cimatti}, {Elbaz}, {Frayer}, {Renzini}, {Pope}, {Alexander}, {Bauer},
  {Giavalisco}, {Huynh}, {Kurk}, \& {Mignoli}}]{Daddi07}
{Daddi}, E., {Dickinson}, M., {Morrison}, G., {et~al.} 2007, \apj, 670, 156

\bibitem[{{Dey} {et~al.}(2019){Dey}, {Schlegel}, {Lang}, {Blum}, {Burleigh},
  {Fan}, {Findlay}, {Finkbeiner}, {Herrera}, {Juneau}, {Landriau}, {Levi},
  {McGreer}, {Meisner}, {Myers}, {Moustakas}, {Nugent}, {Patej}, {Schlafly},
  {Walker}, {Valdes}, {Weaver}, {Y{\`e}che}, {Zou}, {Zhou}, {Abareshi},
  {Abbott}, {Abolfathi}, {Aguilera}, {Alam}, {Allen}, {Alvarez}, {Annis},
  {Ansarinejad}, {Aubert}, {Beechert}, {Bell}, {BenZvi}, {Beutler}, {Bielby},
  {Bolton}, {Brice{\~n}o}, {Buckley-Geer}, {Butler}, {Calamida}, {Carlberg},
  {Carter}, {Casas}, {Castander}, {Choi}, {Comparat}, {Cukanovaite}, {Delubac},
  {DeVries}, {Dey}, {Dhungana}, {Dickinson}, {Ding}, {Donaldson}, {Duan},
  {Duckworth}, {Eftekharzadeh}, {Eisenstein}, {Etourneau}, {Fagrelius},
  {Farihi}, {Fitzpatrick}, {Font-Ribera}, {Fulmer}, {G{\"a}nsicke},
  {Gaztanaga}, {George}, {Gerdes}, {Gontcho}, {Gorgoni}, {Green}, {Guy},
  {Harmer}, {Hernandez}, {Honscheid}, {Huang}, {James}, {Jannuzi}, {Jiang},
  {Joyce}, {Karcher}, {Karkar}, {Kehoe}, {Kneib}, {Kueter-Young}, {Lan},
  {Lauer}, {Le Guillou}, {Le Van Suu}, {Lee}, {Lesser}, {Perreault Levasseur},
  {Li}, {Mann}, {Marshall}, {Mart{\'\i}nez-V{\'a}zquez}, {Martini}, {du Mas des
  Bourboux}, {McManus}, {Meier}, {M{\'e}nard}, {Metcalfe},
  {Mu{\~n}oz-Guti{\'e}rrez}, {Najita}, {Napier}, {Narayan}, {Newman}, {Nie},
  {Nord}, {Norman}, {Olsen}, {Paat}, {Palanque-Delabrouille}, {Peng},
  {Poppett}, {Poremba}, {Prakash}, {Rabinowitz}, {Raichoor}, {Rezaie},
  {Robertson}, {Roe}, {Ross}, {Ross}, {Rudnick}, {Safonova}, {Saha},
  {S{\'a}nchez}, {Savary}, {Schweiker}, {Scott}, {Seo}, {Shan}, {Silva},
  {Slepian}, {Soto}, {Sprayberry}, {Staten}, {Stillman}, {Stupak}, {Summers},
  {Sien Tie}, {Tirado}, {Vargas-Maga{\~n}a}, {Vivas}, {Wechsler}, {Williams},
  {Yang}, {Yang}, {Yapici}, {Zaritsky}, {Zenteno}, {Zhang}, {Zhang}, {Zhou}, \&
  {Zhou}}]{Dey19}
{Dey}, A., {Schlegel}, D.~J., {Lang}, D., {et~al.} 2019, \aj, 157, 168

\bibitem[{{Elmegreen} {et~al.}(2012){Elmegreen}, {Zhang}, \&
  {Hunter}}]{Elmegreen12}
{Elmegreen}, B.~G., {Zhang}, H.-X., \& {Hunter}, D.~A. 2012, \apj, 747, 105

\bibitem[{{Faber}(1973)}]{Faber73}
{Faber}, S.~M. 1973, \apj, 179, 423

\bibitem[{{Ferrarese} {et~al.}(2006){Ferrarese}, {C{\^o}t{\'e}}, {Jord{\'a}n},
  {Peng}, {Blakeslee}, {Piatek}, {Mei}, {Merritt}, {Milosavljevi{\'c}},
  {Tonry}, \& {West}}]{Ferrarese06}
{Ferrarese}, L., {C{\^o}t{\'e}}, P., {Jord{\'a}n}, A., {et~al.} 2006, \apjs,
  164, 334

\bibitem[{{Geha} {et~al.}(2005){Geha}, {Guhathakurta}, \& {van der
  Marel}}]{Geha05}
{Geha}, M., {Guhathakurta}, P., \& {van der Marel}, R.~P. 2005, \aj, 129, 2617

\bibitem[{{Graham} \& {Guzm{\'a}n}(2003)}]{Graham03}
{Graham}, A.~W., \& {Guzm{\'a}n}, R. 2003, \aj, 125, 2936

\bibitem[{{Huxor} {et~al.}(2013){Huxor}, {Phillipps}, \& {Price}}]{Huxor13}
{Huxor}, A.~P., {Phillipps}, S., \& {Price}, J. 2013, \mnras, 430, 1956

\bibitem[{{Huxor} {et~al.}(2011){Huxor}, {Phillipps}, {Price}, \&
  {Harniman}}]{Huxor11}
{Huxor}, A.~P., {Phillipps}, S., {Price}, J., \& {Harniman}, R. 2011, \mnras,
  414, 3557

\bibitem[{{Janz} \& {Lisker}(2008)}]{Janz08}
{Janz}, J., \& {Lisker}, T. 2008, \apjl, 689, L25

\bibitem[{{Jedrzejewski}(1987)}]{Jedrzejewski87}
{Jedrzejewski}, R.~I. 1987, \mnras, 226, 747

\bibitem[{{Kennicutt}(1998)}]{Kennicutt98}
{Kennicutt}, Jr., R.~C. 1998, \araa, 36, 189

\bibitem[{{Kim} {et~al.}(2020){Kim}, {Jeong}, {Rey}, {Lee}, {Lee}, {Joo}, \&
  {Kim}}]{Kim20}
{Kim}, S., {Jeong}, H., {Rey}, S.-C., {et~al.} 2020, \apj, 903, 65

\bibitem[{{Kormendy}(1985)}]{Kormendy85}
{Kormendy}, J. 1985, \apj, 295, 73

\bibitem[{{Kormendy} {et~al.}(2009){Kormendy}, {Fisher}, {Cornell}, \&
  {Bender}}]{Kormendy09}
{Kormendy}, J., {Fisher}, D.~B., {Cornell}, M.~E., \& {Bender}, R. 2009, \apjs,
  182, 216

\bibitem[{{Kravtsov} {et~al.}(2018){Kravtsov}, {Vikhlinin}, \&
  {Meshcheryakov}}]{Kravtsov18}
{Kravtsov}, A.~V., {Vikhlinin}, A.~A., \& {Meshcheryakov}, A.~V. 2018,
  Astronomy Letters, 44, 8

\bibitem[{{Lee} {et~al.}(2009){Lee}, {Kennicutt}, {Funes}, {Sakai}, \&
  {Akiyama}}]{Lee09}
{Lee}, J.~C., {Kennicutt}, Robert~C., J., {Funes}, S.~J. J.~G., {Sakai}, S., \&
  {Akiyama}, S. 2009, \apj, 692, 1305

\bibitem[{{Leitherer} {et~al.}(1999){Leitherer}, {Schaerer}, {Goldader},
  {Delgado}, {Robert}, {Kune}, {de Mello}, {Devost}, \&
  {Heckman}}]{Leitherer99}
{Leitherer}, C., {Schaerer}, D., {Goldader}, J.~D., {et~al.} 1999, \apjs, 123,
  3

\bibitem[{{Marino} {et~al.}(2013){Marino}, {Rosales-Ortega}, {S{\'a}nchez},
  {Gil de Paz}, {V{\'{\i}}lchez}, {Miralles-Caballero}, {Kehrig},
  {P{\'e}rez-Montero}, {Stanishev}, {Iglesias-P{\'a}ramo}, {D{\'{\i}}az},
  {Castillo-Morales}, {Kennicutt}, {L{\'o}pez-S{\'a}nchez}, {Galbany},
  {Garc{\'{\i}}a-Benito}, {Mast}, {Mendez-Abreu}, {Monreal-Ibero}, {Husemann},
  {Walcher}, {Garc{\'{\i}}a-Lorenzo}, {Masegosa}, {Del Olmo Orozco},
  {Mour{\~a}o}, {Ziegler}, {Moll{\'a}}, {Papaderos},
  {S{\'a}nchez-Bl{\'a}zquez}, {Gonz{\'a}lez Delgado}, {Falc{\'o}n-Barroso},
  {Roth}, {van de Ven}, \& {Califa Team}}]{Marino13}
{Marino}, R.~A., {Rosales-Ortega}, F.~F., {S{\'a}nchez}, S.~F., {et~al.} 2013,
  \aap, 559, A114

\bibitem[{{Mieske} {et~al.}(2002){Mieske}, {Hilker}, \& {Infante}}]{Mieske02}
{Mieske}, S., {Hilker}, M., \& {Infante}, L. 2002, \aap, 383, 823

\bibitem[{{Mihos} \& {Hernquist}(1994)}]{Mihos94}
{Mihos}, J.~C., \& {Hernquist}, L. 1994, \apjl, 437, L47

\bibitem[{{Naab} \& {Ostriker}(2009)}]{Naab09}
{Naab}, T., \& {Ostriker}, J.~P. 2009, \apj, 690, 1452

\bibitem[{{Nusser}(2005)}]{Nusser05}
{Nusser}, A. 2005, \mnras, 361, 977

\bibitem[{{Nusser} \& {Silk}(2022)}]{Nusser22}
{Nusser}, A., \& {Silk}, J. 2022, \mnras, 509, 2979

\bibitem[{{Papaderos} {et~al.}(1996){Papaderos}, {Loose}, {Thuan}, \&
  {Fricke}}]{Papaderos96}
{Papaderos}, P., {Loose}, H.-H., {Thuan}, T.~X., \& {Fricke}, K.~J. 1996,
  \aaps, 120, 207

\bibitem[{{Paudel} {et~al.}(2016){Paudel}, {Hilker}, {Ree}, \&
  {Kim}}]{Paudel16}
{Paudel}, S., {Hilker}, M., {Ree}, C.~H., \& {Kim}, M. 2016, \apjl, 820, L19

\bibitem[{{Paudel} {et~al.}(2014){Paudel}, {Lisker}, {Hansson}, \&
  {Huxor}}]{Paudel14}
{Paudel}, S., {Lisker}, T., {Hansson}, K.~S.~A., \& {Huxor}, A.~P. 2014,
  \mnras, 443, 446

\bibitem[{{Paudel} {et~al.}(2018){Paudel}, {Smith}, {Yoon},
  {Calder{\'o}n-Castillo}, \& {Duc}}]{Paudel18}
{Paudel}, S., {Smith}, R., {Yoon}, S.~J., {Calder{\'o}n-Castillo}, P., \&
  {Duc}, P.-A. 2018, \apjs, 237, 36

\bibitem[{{Penny} {et~al.}(2012){Penny}, {Pimbblet}, {Conselice}, {Brown},
  {Gr{\"u}tzbauch}, \& {Floyd}}]{Penny12}
{Penny}, S.~J., {Pimbblet}, K.~A., {Conselice}, C.~J., {et~al.} 2012, \apjl,
  758, L32

\bibitem[{{Petrosian}(1976)}]{Petrosian76}
{Petrosian}, V. 1976, \apjl, 209, L1

\bibitem[{{Rey} {et~al.}(2021){Rey}, {Oh}, \& {Kim}}]{Rey21}
{Rey}, S.-C., {Oh}, K., \& {Kim}, S. 2021, \apjl, 917, L9

\bibitem[{{Rich} {et~al.}(2012){Rich}, {Collins}, {Black}, {Longstaff}, {Koch},
  {Benson}, \& {Reitzel}}]{Rich12}
{Rich}, R.~M., {Collins}, M.~L.~M., {Black}, C.~M., {et~al.} 2012, \nat, 482,
  192

\bibitem[{{Sarzi} {et~al.}(2006){Sarzi}, {Falc{\'o}n-Barroso}, {Davies},
  {Bacon}, {Bureau}, {Cappellari}, {de Zeeuw}, {Emsellem}, {Fathi},
  {Krajnovi{\'c}}, {Kuntschner}, {McDermid}, \& {Peletier}}]{Sarzi06}
{Sarzi}, M., {Falc{\'o}n-Barroso}, J., {Davies}, R.~L., {et~al.} 2006, \mnras,
  366, 1151

\bibitem[{{Sersic}(1968)}]{Sersic68}
{Sersic}, J.~L. 1968, {Atlas de galaxias australes}, ed. {Sersic, J.~L.}

\bibitem[{{Stierwalt} {et~al.}(2015){Stierwalt}, {Besla}, {Patton}, {Johnson},
  {Kallivayalil}, {Putman}, {Privon}, \& {Ross}}]{Stierwalt15}
{Stierwalt}, S., {Besla}, G., {Patton}, D., {et~al.} 2015, \apj, 805, 2

\bibitem[{{Sung} {et~al.}(1998){Sung}, {Han}, {Ryden}, {Chun}, \&
  {Kim}}]{Sung98}
{Sung}, E.-C., {Han}, C., {Ryden}, B.~S., {Chun}, M.-S., \& {Kim}, H.-I. 1998,
  \apj, 499, 140

\bibitem[{{Toloba} {et~al.}(2014){Toloba}, {Guhathakurta}, {van de Ven},
  {Boissier}, {Boselli}, {den Brok}, {Falc{\'o}n-Barroso}, {Hensler}, {Janz},
  {Laurikainen}, {Lisker}, {Paudel}, {Peletier}, {Ry{\'s}}, \&
  {Salo}}]{Toloba14}
{Toloba}, E., {Guhathakurta}, P., {van de Ven}, G., {et~al.} 2014, \apj, 783,
  120

\bibitem[{{Toomre} \& {Toomre}(1972)}]{Toomre72}
{Toomre}, A., \& {Toomre}, J. 1972, \apj, 178, 623

\bibitem[{{Tremonti} {et~al.}(2004){Tremonti}, {Heckman}, {Kauffmann},
  {Brinchmann}, {Charlot}, {White}, {Seibert}, {Peng}, {Schlegel}, {Uomoto},
  {Fukugita}, \& {Brinkmann}}]{Tremonti04}
{Tremonti}, C.~A., {Heckman}, T.~M., {Kauffmann}, G., {et~al.} 2004, \apj, 613,
  898

\bibitem[{{Wellons} {et~al.}(2015){Wellons}, {Torrey}, {Ma}, {Rodriguez-Gomez},
  {Vogelsberger}, {Kriek}, {van Dokkum}, {Nelson}, {Genel}, {Pillepich},
  {Springel}, {Sijacki}, {Snyder}, {Nelson}, {Sales}, \&
  {Hernquist}}]{Wellons15}
{Wellons}, S., {Torrey}, P., {Ma}, C.-P., {et~al.} 2015, \mnras, 449, 361

\bibitem[{{Wellons} {et~al.}(2016){Wellons}, {Torrey}, {Ma}, {Rodriguez-Gomez},
  {Pillepich}, {Nelson}, {Genel}, {Vogelsberger}, \& {Hernquist}}]{Wellons16}
{Wellons}, S., {Torrey}, P., {Ma}, C.-P., {et~al.} 2016, \mnras, 456, 1030

\bibitem[{{Zhang} {et~al.}(2024){Zhang}, {Zhu}, {Jiang}, {Cheng}, {Wang},
  {Wang}, {Xu}, {Liu}, {Yu}, {Qian}, {Yu}, {Ai}, {Jing}, {Xu}, {Liu}, {Guan},
  {Sun}, {Yang}, {Huang}, {Hao}, \& {FAST Collaboration}}]{Zhang24}
{Zhang}, C.-P., {Zhu}, M., {Jiang}, P., {et~al.} 2024, Science China Physics,
  Mechanics, and Astronomy, 67, 219511

\bibitem[{{Zhang} {et~al.}(2020{\natexlab{a}}){Zhang}, {Smith}, {Oh}, {Paudel},
  {Duc}, {Boselli}, {C{\^o}t{\'e}}, {Ferrarese}, {Gao}, {Hunter}, {Puzia},
  {Peng}, {Rong}, {Shin}, \& {Zhao}}]{Zhang20a}
{Zhang}, H.-X., {Smith}, R., {Oh}, S.-H., {et~al.} 2020{\natexlab{a}}, \apj,
  900, 152

\bibitem[{{Zhang} {et~al.}(2020{\natexlab{b}}){Zhang}, {Paudel}, {Smith},
  {Duc}, {Puzia}, {Peng}, {C{\^o}te}, {Ferrarese}, {Boselli}, {Wang}, \&
  {Oh}}]{Zhang20b}
{Zhang}, H.-X., {Paudel}, S., {Smith}, R., {et~al.} 2020{\natexlab{b}}, \apjl,
  891, L23

\bibitem[{{Zhang} \& {Bell}(2017)}]{Zhang17}
{Zhang}, Y., \& {Bell}, E.~F. 2017, \apjl, 835, L2

\end{thebibliography}


\end{document}